\documentclass[aps,preprint]{revtex4-1} 
\usepackage{graphicx}
\usepackage{amssymb}
\usepackage{epstopdf}
\usepackage{amsmath}
\usepackage{bm}

\newcommand{\half}{\frac{1}{2}}

\newcommand{\dd}{\partial}

\newcommand\refeq[1]{(\ref{#1})}
\newcommand\reffig[1]{Figure~\ref{#1}}
\newcommand\of[1]{\left( #1 \right)}

\newcommand\vecbf[1]{{\bf #1}}

\newcommand\mat[1]{{\mathbb #1}}

\begin{document}

\title{The Schr\"odinger-Newton System with Self-Field Coupling}

\author{J. Franklin}
\email{jfrankli@reed.edu}
\author{Y. Guo}
\author{A. McNutt}
\author{A. Morgan}

\affiliation{Department of Physics, Reed College, Portland, Oregon 97202,
USA}

\begin{abstract}
We study the Schr\"odinger-Newton system of equations with the addition of gravitational field energy sourcing -- such additional nonlinearity is to be expected from a theory of gravity (like general relativity), and its appearance in this simplified scalar setting (one of Einstein's precursors to general relativity) leads to significant changes in the spectrum of the self-gravitating theory.   Using an iterative technique, we compare the mass dependence of the ground state energies of both Schr\"odinger-Newton and the new, self-sourced system and find that they are dramatically different.  The Bohr method approach from old quantization provides a qualitative description of the difference, which comes from the additional nonlinearity introduced in the self-sourced case.  In addition to comparison of ground state energies, we calculate the transition energy between the ground state and first excited state to compare emission frequencies between Schr\"odinger-Newton and the self-coupled scalar case.
\end{abstract}

\maketitle

\section{Introduction}

The Schr\"odinger-Newton (SN) system of equations has been studied in various contexts since its introduction in~\cite{Bon} as a model for self-gravitating (quantum) particles.  The single-particle formulation of the problem treats the central body's quantum mechanical distribution as a mass density sourcing the gravitational field (whose influence affects the central body):
\begin{equation}\label{NSet}
\begin{aligned}
 i \, \hbar \, \frac{\dd \Psi(\vecbf r,t)}{\dd t} &= -\frac{\hbar^2}{2 \, m} \nabla^2 \Psi(\vecbf r,t) + m\, \Phi(\vecbf r, t) \, \Psi(\vecbf r,t) \\
\nabla^2 \Phi(\vecbf r, t) &= 4 \, \pi \, G \, m \, \Psi^*(\vecbf r, t) \, \Psi(\vecbf r, t)
\end{aligned}
\end{equation}
where $\Psi(\vecbf r, t)$ is the wave function associated with the particle of mass $m$ moving under the influence of the potential energy $\Phi(\vecbf r, t)$.

Regardless of its motivation (and there are a variety of motivations for studying this set -- some of these are described in~\cite{Carlip,Giulini2} and references  therein -- we side-step that discussion in the current work, focusing instead on a comparative study of the solutions to~\refeq{NSet} and our proposed modification), much is known about the solutions in both the time-independent~\cite{Moroz, Tod} and time-dependent form~\cite{HarrisonN,Meter, Manfredi} (see also the dissertations~\cite{Harrison,Salzman} for additional background and review).

We propose to augment the gravitational sourcing in~\refeq{NSet} to include the energy density present in the gravitational field itself.  The motivation for doing this comes originally from special relativity and the universal coupling of gravity -- if mass density acts as a source, then so can energy density, and the energy density of the gravitational field is available as a source prior to any additional external sources (like the energy density associated with electromagnetic fields, for example).  The governing equation for $\Phi$, the gravitational potential energy, is then~\cite{Einstein}:
\begin{equation}\label{EG}
\nabla^2 \Phi(\vecbf r) = \frac{4 \, \pi \, G }{c^2} \,  \rho(\vecbf r)\, \Phi(\vecbf r) + \frac{1}{2 \, \Phi(\vecbf r)} \, \nabla \Phi(\vecbf r) \cdot \nabla \Phi(\vecbf r).
\end{equation}
This is the field equation for a scalar $\Phi(\vecbf r)$ that is sourced by a distribution of mass given by the mass density $\rho(\vecbf r)$ and its own energy density (represented by the second term).  This field equation was developed originally by Einstein~\cite{Einstein}, and a (special) relativistic version appeared in~\cite{FandN} -- more recently, the self-coupled case was re-derived in~\cite{GiuliniSC}.


If one were to similarly couple a second-rank field theory to itself by making the field's stress tensor a source -- motivated by the universal nature of gravity -- one ends up with general relativity (GR)~\cite{Deser} (and for a review of that process~\cite{Franklin}) together with its geometric interpretation.  The sourcing in~\refeq{EG} is a scalar version of that self-coupling (and was worked out in that context in~\cite{DandH}).  Our fundamental question will be how this relatively simple nonlinearity, showing up in the gravitational field equation \footnote{Any theory of gravity should include some sort of nonlinearity due to self-sourcing, so it is important to introduce such terms, even at the toy scalar level.} changes the spectrum of the resulting Schr\"odinger-coupled system.

We'll start with a review of the self-coupling that leads to the (static) field theory~\refeq{EG}, then develop the spherically symmetric vacuum solution of~\refeq{EG}, and use that to estimate the ground state energy as a function of particle mass.  Then we will introduce a numerical method to calculate the ground state energy of both SN (to test correct numerical behavior) and the new self-coupled gravitational field, and compare the ground state energy to the predicted form.  For both the ground state and the first excited state, the Bohr model estimates provide qualitatively correct predictions.  By calculating the energy difference between the first excited state and ground state of the self-coupled gravity, we can compare the transition energy emission of this theory with the one predicted by SN.  Finally, we calculate the total energy of $N$ self-gravitating bosons using the self-coupled form, and compare that with the total energy as calculated using SN in~\cite{Bon}.


\section{Motivation}

The static gravitational field theory given by~\refeq{EG} was first proposed by Einstein {\it en route} to general relativity.  We will generate the field equation here (a recent discussion of this derivation is in~\cite{FranklinAJP}) -- the issue that motivates its development is the lack of self-coupling of the original field equation of Newtonian gravity: $\nabla^2 \Phi = 4 \, \pi \, G \, \rho$ describes a field $\Phi$ sourced by a mass density $\rho$, but this Poisson form lacks the energy density source that comes from $\Phi$ itself.

Let's start by calculating the energy density associated with a field $\Phi$ that comes from Newtonian gravity (so that it satisfies the Poisson equation: $\nabla^2 \Phi = 4 \, \pi \, G \, \rho$).  Proceeding as usual, the work done in building a distribution of mass, $\rho$, is given by:
\begin{equation}
W = \half \, \int_{\hbox{\tiny{all space}}} \rho \, \Phi \, d\tau,
\end{equation}
then using the Poisson equation to eliminate $\rho$ and integrating by parts gives
\begin{equation}
W = \int \of{ -\frac{1}{8 \, \pi \, G} \, \nabla\Phi \cdot \nabla \Phi } \, d\tau,
\end{equation}
and we would call the integrand the energy density of the field:
\begin{equation}\label{ufirstpass}
u_\Phi = -\frac{1}{8 \, \pi \, G} \, \nabla \Phi \cdot \nabla \Phi.
\end{equation}
According to special relativity, we should use an associated effective mass density $\rho_\Phi = u_\Phi/c^2$ as a source in the field equation for $\Phi$.  That is the nonlinearity promised by a theory of gravity -- the field's self-energy must act as a source.  So we are tempted to start with:
\begin{equation}\label{PetersP}
\nabla^2 \Phi = 4 \, \pi \, G \, \of{\rho + \rho_\Phi} = 4 \, \pi \, G \, \rho - \frac{1}{2 \, c^2} \nabla \Phi \cdot \nabla\Phi,
\end{equation}
and this form was considered in~\cite{Peters}.  The problem is that the new field equation~\refeq{PetersP} leads to an energy density (obtained as above, but with the new field equation instead of the Poisson equation) that is {\it not}~\refeq{ufirstpass}, and so the self-coupling is not consistent.

We'll now generate the correct self-coupled theory using the machinery of field theory, where the Hamiltonian density is the energy density (for a static, free field).
Suppose we start with a free field theory that has Hamiltonian $\mathcal{H} = \frac{f(\Phi)}{8 \, \pi \, G} \, \nabla \Phi \cdot \nabla \Phi$, i.e.\ we augment the usual scalar Hamiltonian with a function of $\Phi$, $f(\Phi)$, that we will fix by demanding that it lead to a source term that looks like $\rho_{\mathcal{H}}= \mathcal{H}/c^2$.  The field equation for this $\mathcal{H}$ is
\begin{equation}\label{protP}
\frac{f(\Phi)}{4 \, \pi \, G} \, \nabla^2 \Phi + \frac{f'(\Phi)}{8 \, \pi \, G} \, \nabla \Phi \cdot \nabla \Phi = 0,
\end{equation}
or
\begin{equation}
\nabla^2 \Phi = -\frac{f'(\Phi)}{2 \, f(\Phi)} \, \nabla\Phi \cdot \nabla \Phi.
\end{equation}
It is the right-hand side of this equation that we would like to set equal to $4 \, \pi \, G\, \of{\mathcal{H}/c^2} = \frac{f(\Phi)}{2 \, c^2} \, \nabla \Phi \cdot \nabla \Phi$, giving us the self-consistent energy density source, and when we do that, we get an ODE for $f(\Phi)$:
\begin{equation}
\frac{f(\Phi)}{2\, c^2} =  -\frac{f'(\Phi)}{2 \, f(\Phi)}  \longrightarrow f(\Phi) = \frac{c^2}{\Phi}.
\end{equation}

Our final free Hamiltonian is: $\mathcal H = \frac{c^2}{8 \, \pi \, G \, \Phi} \, \nabla \Phi \cdot \nabla \Phi$, and this is the energy density of the field in the theory, given also in~\cite{Einstein}.  Using this solution for $f(\Phi)$ gives the field equation:
\begin{equation}
\nabla^2 \Phi = \frac{1}{2 \, \Phi} \, \nabla\Phi \cdot \nabla \Phi,
\end{equation}
in vacuum, and we recover~\refeq{EG} when we introduce massive sources (that would put a factor of $\rho$ on the right-hand side of~\refeq{protP}).  When computing solutions to this self-consistent, self-coupled form of gravity, it is worth noting that the field equation~\refeq{EG} can be written linearly in $\sqrt{\Phi}$, where it reads (again, from~\cite{Einstein}):
\begin{equation}\label{Einsimple}
\nabla^2 \of{ \sqrt{\Phi}} = \frac{2 \, \pi \, G}{c^2} \, \rho \, \of{\sqrt{\Phi}}.
\end{equation}
In order to recover Newtonian gravity, we must take $\Phi = c^2 + \Phi_N$, then inserting this into~\refeq{Einsimple} (or~\refeq{EG}) and collecting in powers of $c$ gives back $\nabla^2 \Phi_N = 4 \, \pi \, G \, \rho$ to zeroth order, a requirement of the weak-field limit.

\section{Point Source Solution}

We'll start by looking at solutions to the field equation:
\begin{equation}\label{Pfield}
\nabla^2 \Phi = \frac{4 \, \pi \, G \, \rho}{c^2} \, \Phi + \frac{1}{2 \, \Phi} \, \nabla \Phi \cdot \nabla \Phi,
\end{equation}
in regions where $\rho = 0$ (so we are in vacuum) and $\Phi(\vecbf r) = \Phi(r)$.  Such a solution would be appropriate for a spherically symmetric source of mass $m$ localized near the origin.
Under our assumptions, the field equation reduces to
\begin{equation}
\left( r \, \Phi(r) \right)'' = \frac{r}{2 \, \Phi(r)} \, \left( \Phi'(r) \right)^2
\end{equation}
with primes denoting $r$-derivatives.  The solution to this equation comes with two integration constants, $\alpha$ and $\beta$:
\begin{equation}
\Phi(r) = \beta \, \left[ 4 \, \alpha^2 - \frac{4 \, \alpha}{r} + \frac{1}{r^2} \right].
\end{equation}
If we ask that $\Phi(r)$ look like a Newtonian point source (of mass $m$) as $r$ approaches spatial infinity, then we can fix $\alpha$
\begin{equation}\label{prebchoice}
\Phi(r) = -\frac{G \, m}{r} + \frac{\beta}{r^2} + \frac{G^2 \, m^2}{4 \, \beta}.
\end{equation}
To recover the Poisson form of Newtonian gravity for $\Phi$ small (compared to $c^2$), we must have $\Phi(\infty) = c^2$ as a boundary condition, and that sets the constant $\beta$.  Our final spherically symmetric vacuum solution looks like
\begin{equation}\label{ppSC}
\Phi(r) = -\frac{G \, m}{r} + \frac{G^2 \, m^2}{4 \, c^2 \, r^2} + c^2.
\end{equation}
The energy density of this field is, from $\mathcal H = \frac{c^2}{8 \, \pi \, G \, \Phi} \, \nabla \Phi \cdot \nabla \Phi$,
\begin{equation}
\mathcal{H} = \frac{G \, m^2}{8 \, \pi \, r^4},
\end{equation}
which is everywhere positive.

We are interested in the vacuum solution because it is the one that is relevant to the Bohr approach to estimating quantum mechanical energies.  But it is easy to compute (especially from~\refeq{Einsimple}) the solution in cases other than vacuum -- for example, if we had a sphere of radius $R$ with constant density $\rho_0$ inside it, then the interior solution to~\refeq{EG} is just:
\begin{equation}
\Phi_i(r) = \of{\frac{A \, \sinh\of{r/r_0}}{r}}^2 \, \, \, \, \, \, \, \, \, \, r_0 \equiv \frac{c}{\sqrt{2 \, \pi \, G \, \rho_0}},
\end{equation}
where $A$ is a constant that we would use to match up to an exterior solution (at $r > R$, for example) and we have chosen the solution that is finite at the origin.

\section{Scaling of the Ground State}

For the SN system, with constants $G$, $\hbar$, and $m$, there is only one way to make an energy
\begin{equation}
E \sim \frac{G^2}{\hbar^2} \, m^5,
\end{equation}
so we expect, up to constants out front, that the energy spectrum of~\refeq{NSet} scales like $m^5$ (as indeed it does~\cite{Harrison, Bernstein}).

With the introduction of $c$ appearing in~\refeq{EG}, we can form the Planck mass: $M_p = \sqrt{\frac{\hbar \, c}{G}}$, and this means that any power of $m$ could appear in the energy spectrum of the self-coupled system (where we expect $E \sim \frac{G^2}{\hbar^2} \, m^q \, M_p^s$ with $q + s  = 5$, but otherwise unconstrained).   We'd like a way to estimate relevant combinations of $M_p$ and $m$ that might appear in our new spectrum.  To that end, we will use the Bohr model (originally for hydrogen, of course, but applied in this gravitational setting) with potential given by~\refeq{ppSC} -- the idea is that if the ground state is localized close to the origin, then far away, the potential associated with that ground state should go roughly like~\refeq{ppSC}, and so the spectrum of the spherically symmetric vacuum solution could provide some relevant approximate information.

To start, we'll apply the Bohr method to $\Phi(r) = -\frac{G \, m}{r} + c^2$, just the Newtonian point particle potential (with an offset at spatial infinity so as to match~\refeq{ppSC}).  According to the rules of old quantization (for circular orbits, updated to elliptical orbits in the Wilson-Sommerfeld formulation), we start with 
the total energy for a particle moving in a circle of radius $r$ (so that $v = \sqrt{G \, m/r}$):
\begin{equation}
E= m\, c^2 -\frac{1}{2} \, \frac{G \, m^2}{r}.
\end{equation}
Next, we assume angular momentum is quantized: $L = m \, v \, r = n \, \hbar$ for integer $n$, giving us a value for the radius: $r = \frac{n^2\, \hbar^2}{G \, m^3}$ -- then using this radius in $E$ we get a discrete set of energies:
\begin{equation}
E_n = m \, c^2 - \half \, \frac{G^2 \, m^5}{n^2\,  \hbar^2}.
\end{equation}
The ground state corresponds to $n=1$:
\begin{equation}\label{BFitSN}
E_1 = m \, c^2 -\frac{1}{2} \,  \frac{G^2}{\hbar^2} \, m^5 = m\, c^2  -\frac{1}{2} \, \frac{c^2}{M_p^4}  m^5,
\end{equation}
which is what we expect, namely a linear (in $m$) offset (associated with the shift at spatial infinity) and $m^5$ scaling.

Performing the same procedure for the ``point potential" in~\refeq{ppSC} gives
\begin{equation}\label{BohrSC}
E_n = \frac{2 \, m \, c^2 \, M_p^4 \, n^2}{m^4 + 2 \, M_p^4 \, n^2},
\end{equation}
with ground state energy:
\begin{equation}\label{BFitSCSC}
E_1 = \frac{2 \, m \, c^2 \, M_p^4}{m^4 + 2 \, M_p^4},
\end{equation}
where again, we only care about the mass scaling here -- our estimate cannot predict constant offsets and/or overall constants out front.  We'll come back to those later on.
Note that the expression~\refeq{BFitSCSC} reduces to~\refeq{BFitSN} in the $m \ll M_P$ limit, as it should:
\begin{equation}
 \frac{2 \, m \, c^2 \, M_p^4}{m^4 + 2 \, M_p^4} = m\, c^2  -\frac{1}{2} \, \frac{c^2}{M_p^4}  m^5 + O\of{\of{\frac{m}{M_p}}^8} \, m \, c^2
\end{equation}
for $m$ small.

\section{Numerical Approach}
We'll start by specializing to spherically symmetric solutions, then we'll render  the equations of interest dimensionless, and put them in time-independent form to define the eigenvalue problem of interest.  From there, we'll introduce the iterative finite difference approach that can be applied to either SN or self-coupled gravity to find the ground state energies.

\subsection{Dimensionless Form}
Since we are interested in the ground state energies, we will focus on spherically symmetric solutions to both SN and the modified system.  For SN, we have
\begin{equation}
\begin{aligned}
-\frac{\hbar^2}{2 \, m} \, \frac{1}{r} \, \left(r\, \Psi(r,t) \right)'' + m \, \Phi(r,t) \, \Psi(r,t) &= i \, \hbar \, \frac{\dd \Psi(r,t)}{\dd t} \\
\frac{1}{r} \, \left( r \, \Phi(r,t) \right)'' &= 4 \, \pi \, G \, m \, \Psi(r,t)^* \, \Psi(r,t),
\end{aligned}
\end{equation}
and we can simplify further by setting $P(r,t) \equiv r \, \Psi(r,t)$, then the above becomes
\begin{equation}\label{SNII}
\begin{aligned}
-\frac{\hbar^2}{2 \, m} \, P(r,t)'' + m \, \Phi(r,t) \, P(r,t) &= i \, \hbar \, \frac{\dd P(r,t)}{\dd t} \\
 \left( r \, \Phi(r,t) \right)'' &= \frac{4 \, \pi \, G \, m}{r} \, P(r,t)^* \, P(r,t).
\end{aligned}
\end{equation}
The wave function is normalized to $1$, so that our $P(r,t)$ has:
\begin{equation}\label{normit}
4 \, \pi \, \int_0^\infty  \| P(r,t) \|^2 \, dr = 1.
\end{equation}

The modified theory becomes, under the same assumptions and substitutions,
\begin{equation}\label{SCSCII}
\begin{aligned}
-\frac{\hbar^2}{2 \, m} \, P(r,t)'' + m \, \Phi(r,t) \, P(r,t) &= i \, \hbar \, \frac{\dd P(r,t)}{\dd t} \\
 \left( r \, \Phi(r,t) \right)'' &= \frac{4 \, \pi \, G \, m}{c^2 \, r} \, P(r,t)^* \, P(r,t) \, \Phi(r,t) + \frac{r}{2 \, \Phi(r,t)} \, \left(\Phi(r,t)'\right)^2.
\end{aligned}
\end{equation}

We can render the equations dimensionless by introducing $r = r_0 \, R$, $t = t_0 \, T$ for dimensionless $R$ and $T$, and taking $\Phi = \Phi_0 \, \bar\Phi$, $P = P_0 \, \bar P$, where $\Phi_0$ is a speed$^2$ and $P_0$ has dimension $1/\sqrt{\hbox{length}}$.  Finally, let $m = m_0 \, \bar m$ for Planck mass $m_0 = \sqrt{\frac{\hbar \, c}{G}}$ and set $\bar P(r,t) = e^{-i \, \bar E \, T} \, \bar P(r)$ for dimensionless energy $\bar E = E/E_0$.  Then our pairs take the form of an eigenvalue problem
\begin{equation}\label{SNIII}
\begin{aligned}
-\frac{1}{\bar m} \, \frac{\dd^2 \bar P}{\dd R^2} + \bar m \, \bar \Phi \, \bar P &=  \bar E \, \bar P\\
\frac{\dd^2}{\dd R^2} \left( R \, \bar \Phi \right) &= \frac{\bar m}{R} \, \bar P^* \, \bar P,
\end{aligned}
\end{equation}
and 
\begin{equation}\label{SCSCIII}
\begin{aligned}
-\frac{1}{\bar m}\, \frac{\dd^2 \bar P}{\dd R^2} +  \bar m\, \bar \Phi \, \bar P &=   \bar E \, \bar P \\
\frac{\dd^2}{\dd R^2} \left( R \, \bar \Phi \right) &= \frac{\bar m}{R} \, \bar P^* \, \bar P \, \bar \Phi + \frac{R}{2 \, \bar\Phi} \left( \frac{\dd \bar \Phi}{\dd R}\right)^2
\end{aligned}
\end{equation}
with 
\begin{equation}
r_0 = \frac{\hbar}{\sqrt{2} \, m_0 \, c} \, \, \, \, \, \, \, \, \, \, \, \, t_0 =\frac{\hbar}{m_0 \, c^2}  \, \, \, \, \, \, \, \, \, \, \, \, P_0 =\frac{c}{\sqrt{4 \, \pi \, m_0 \, G}}  \, \, \, \, \, \, \, \, \, \, \, \, \Phi_0 = c^2 \, \, \, \, \, \, \, \, \, \, \, \, E_0 = m_0 \, c^2.
\end{equation}
The normalization of the wave function~\refeq{normit} now reads:
\begin{equation}\label{finalnorm}
\int_0^\infty \bar P^* \, \bar P \, dR = \frac{1}{4 \, \pi \, P_0^2 \, r_0} = \frac{\sqrt{2} \, G \, m_0^2}{\hbar \, c} = \sqrt{2}.
\end{equation}

\subsection{Method}

The numerical method is the same in both cases -- we discretize in $R$ by taking $R_j \equiv j \, \Delta R$, where $j = 0$ is a boundary point (at the origin -- for $\Psi$ finite at the origin, we must have $\bar P = 0$ there), and we take $R_{J+1} = R_\infty$ to be a numerical approximation to infinity, where we again require $\bar P = 0$.  Let $\bar P_j \equiv \bar P(R_j)$ (and $\bar\Phi_j \equiv \bar \Phi(R_j)$), then we can discretize Schr\"odinger's equation using finite differences:
\begin{equation}\label{Schdiff}
-\frac{1}{\bar m} \, \left[ \frac{\bar P_{j+1} - 2 \, \bar P_j + \bar P_{j-1}}{\Delta R^2} \right] + \bar m \, \bar \Phi_j \, \bar P_j = \bar E \, \bar P_j
\end{equation}
for $j = 1 \ldots J$.  We can define the vector $\bar{\vecbf P} \in \mat R^J$ to have entries that are precisely the unknown $\bar P_j$ values (and similarly for the vector $\bar{\bm \Phi}$), and then~\refeq{Schdiff} can be written as a matrix eigenvalue problem in the usual way:
\begin{equation}
\mat D(\bar{\bm \Phi})  \, \bar{\vecbf P} = \bar E \, \bar{\vecbf P}
\end{equation}
with
\begin{equation}\label{DDef}
\mat D(\bar{\bm \Phi}) \dot =\left(\begin{array}{cccc}  \frac{2}{\bar m\, \Delta R^2} + \bar m \, \bar \phi_1 & -\frac{1}{\bar m \, \Delta R^2} & 0 & \ldots \\
-\frac{1}{\bar m \, \Delta R^2} & \frac{2}{\bar m \, \Delta R^2} + \bar m \, \bar \phi^n_2 & -\frac{1}{\bar m\, \Delta R^2} & 0 \\
0 & \ddots & \ddots & \ddots \end{array} \right).
\end{equation}

As a matrix eigenvalue problem, it is relatively easy to construct $\mat D$ and then find the eigenvector associated with the smallest eigenvalue -- that eigenvector is an approximation to the ground state.  Once we have $\bar{\vecbf P}$ in hand, we can construct  the entries of $\bar{\bm \Phi}$ in either the SN or augmented case using Verlet. We can find the values for $\bar\Phi_j$ by working backwards from $j = J$ to $1$ using the recursion
\begin{equation}\label{Verlet}
\bar\Phi_{j-1} = \frac{1}{R_{j-1}} \, \left[ 2 \, R_j \, \bar \Phi_j - R_{j+1} \, \bar{\Phi}_{j+1} + \frac{\bar m}{R_j} \, \| \bar P_j \|^2 \, \Delta R^2\right],
\end{equation}
where we take $\bar\Phi_{j} = 1 - \frac{\sqrt{2} \, \bar m}{R_{j}}$ for $j = J$ and $J+1$ -- since $R_{J+1} = R_\infty$, we want the potential to approximate its value out at spatial infinity, and this is the dimensionless form of $\Phi_j = c^2 - \frac{G \, m}{r_0\, R_j}$.   For the gravitational field equation in~\refeq{SCSCIII}, the analogous Verlet recursion looks like
\begin{equation}\label{VerletSC}
\bar\Phi_{j-1} = \frac{1}{R_{j-1}} \, \biggl[ 2 \, R_j \, \bar \Phi_j - R_{j+1} \, \bar{\Phi}_{j+1} + \frac{\bar m}{R_j} \, \| \bar P_j \|^2 \, \bar{\Phi}_j \, \Delta R^2  + \frac{R_j}{2 \, \bar \Phi_j} \, \left(  \bar \Phi_{j+1} - \bar \Phi_j \right)^2 
\biggr]
\end{equation}
with the same boundary conditions as above.

So, if we had the entires of $\bar{\vecbf P}$, we could construct the entries of $\bar{\bm \Phi}$ in either case, but in order to get  $\bar{\vecbf P}$, we need $\bar{\bm \Phi}$ -- the matrix $\mat D$ in~\refeq{DDef} depends on the values of the potential $\bar{\bm \Phi}$.   We can use an iterative approach (similar to the one in~\cite{Bernstein}) to get around the problem.  Let $\bar P^k_j$ be the $k^{\hbox{\tiny{th}}}$ iteration for $\bar P_j$, then we can construct $\bar \Phi^k_j$ using~\refeq{Verlet}, and update by finding the smallest eigenvalue/vector of $\mat D(\Phi^k)$, so 
\begin{equation}\label{smalleval}
\bar{\vecbf P}^{k+1} = \hbox{smallest eigenvector of $\mat D(\bar{\bm \Phi}^k)$ normalized so that $\sum_{j=1}^n \| \bar P^{k+1}_j\|^2 \, \Delta R = \sqrt{2}$}.
\end{equation}
From this, we can construct $\bar{\bm \Phi}^{k+1}$ and iterate until:
\begin{equation}\label{stopper}
\| \bar{\vecbf P}^{k+1} - \bar{\vecbf P}^k \| \le \epsilon,
\end{equation}
for $\epsilon$, some user-specified tolerance.

\section{Energies}

\subsection{Schr\"odinger Newton}

\begin{figure}[htbp] 
   \centering
   \includegraphics[width=4in]{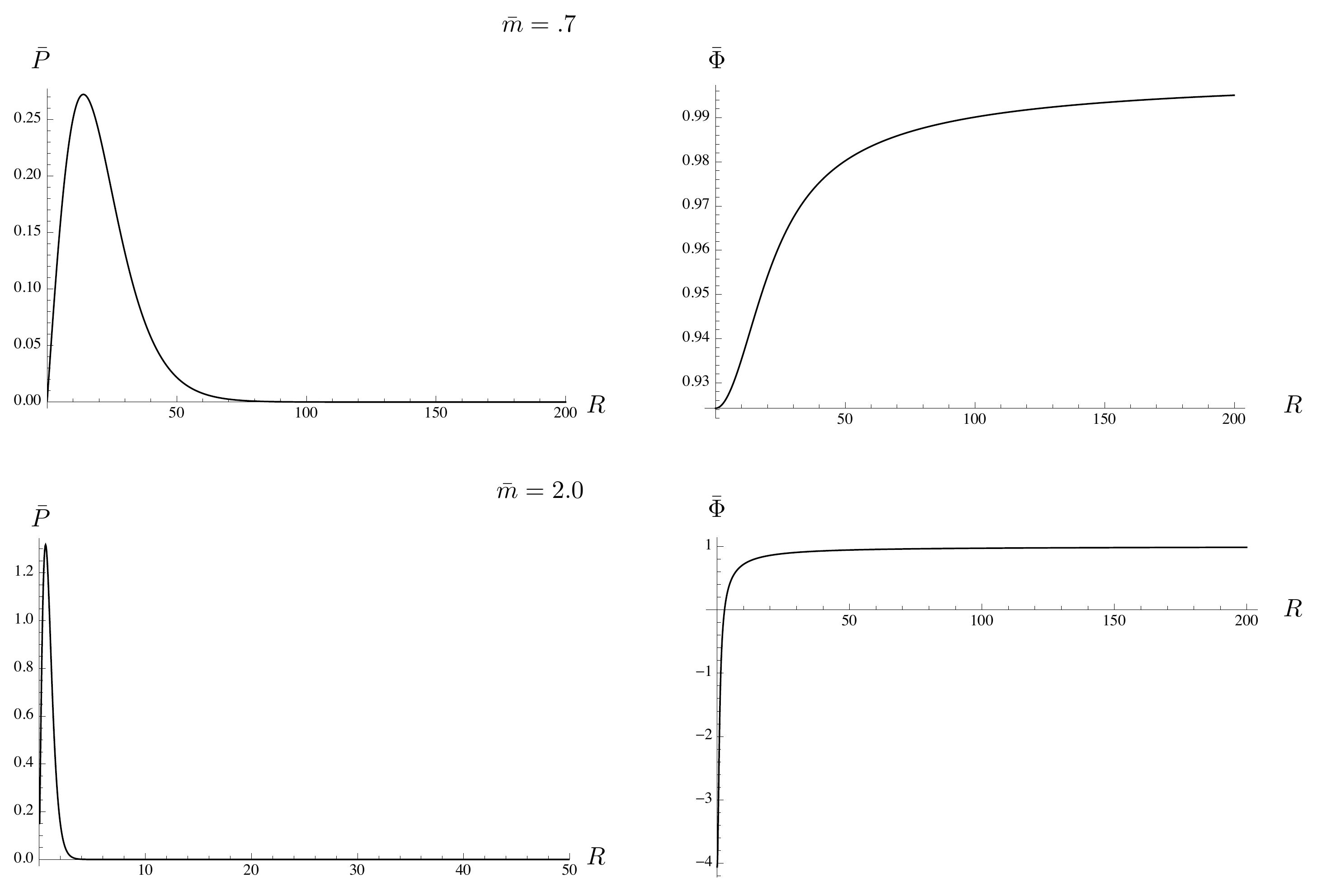} 
   \caption{The numerical solutions for $\bar P$ and $\bar\Phi$ for $\bar m = .7$ (top) and $\bar m = 2$ (bottom) for the SN system.  Here, $R_\infty = 200$ with $\Delta R =200/5001$, and we use $\epsilon = 10^{-8}$ as the tolerance for the iteration (see~\refeq{stopper}).}
   \label{fig:SNPPhi}
\end{figure}

The ground state energies for the SN system were calculated for $\bar m = .7$ to $\bar m = 2$ in steps of $.02$.  In~\reffig{fig:SNPPhi}, we show $\bar P$ and $\bar \Phi$ for the masses $\bar m =.7$ (top) and $\bar m = 2$ (bottom) -- we chose these mass limits because at our value of $R_\infty = 200$ and $\Delta R  = \frac{R_\infty}{5001}$, masses less than $\bar m = .7$ begin to violate our assumption that $\bar P(\infty) = 0$, and masses larger than $\bar m = 2$ get localized to only a few grid points near the origin.  We can expand the mass range by changing $R_\infty$, allowing us to probe smaller masses, and by decreasing $\Delta R$, allowing us to move up in mass.  

\begin{figure}[htbp] 
   \centering
   \includegraphics[width=4in]{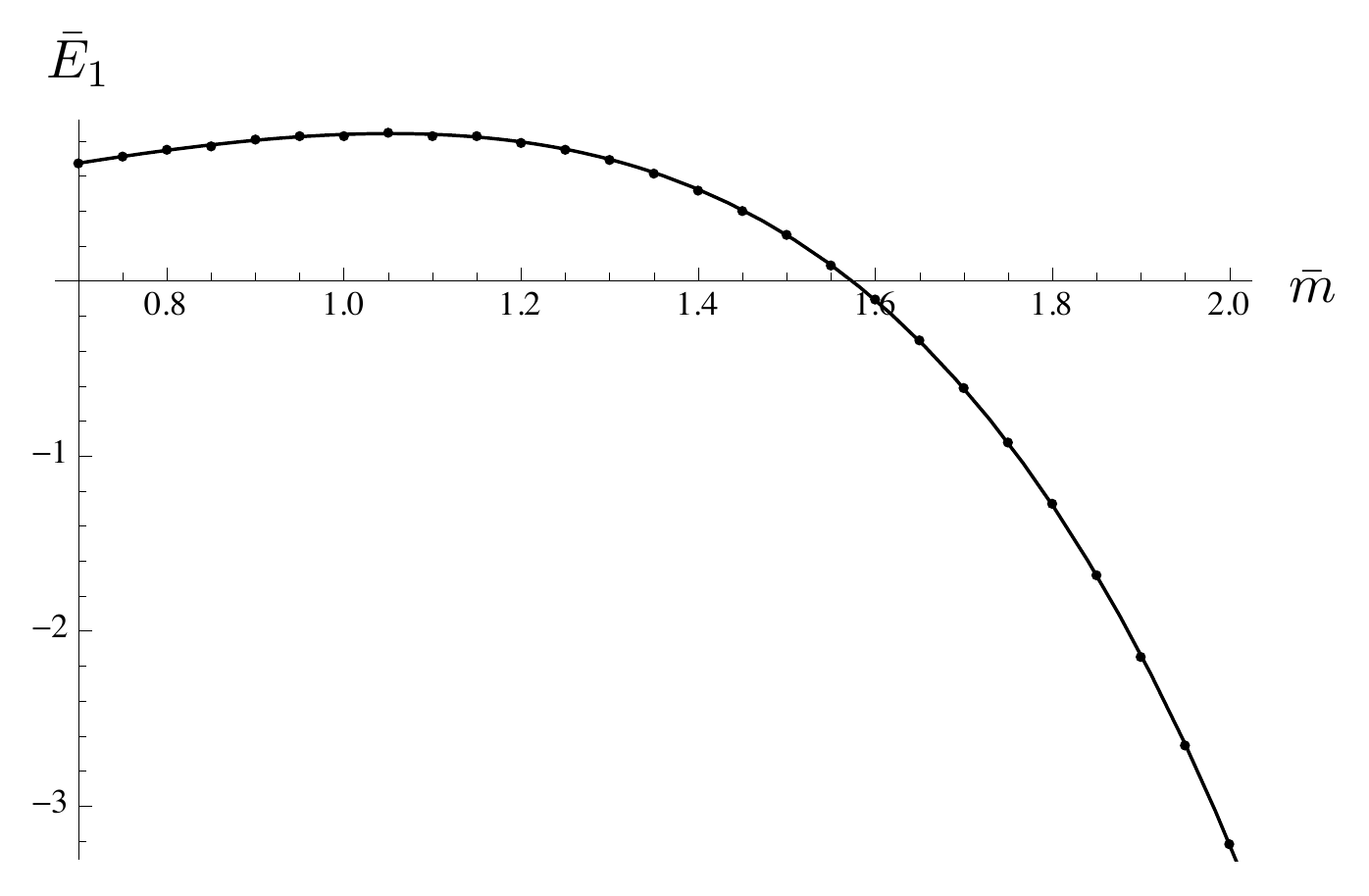} 
   \caption{Ground state energy as a function of mass for the SN pair -- the points are numerically determined using the numerical method described in the previous section, and the curve is the best fit curve, $C(\bar m)$, from~\refeq{BfitSNC}.}
   \label{fig:SNenergy}
\end{figure}

The energies themselves are shown in~\reffig{fig:SNenergy}.  There, the dots are the numerically-determined values -- we fit the dots to a curve of the form $A \, \bar m + B \, \bar m^5$, motivated by~\refeq{BFitSN}, using a nonlinear Levenberg-Marquardt fit of the data.  The curve is:
\begin{equation}\label{BfitSNC}
C(\bar m) = 1.00123\,  \bar m - 0.163181\,  \bar m^5.
\end{equation}
The best fit curve has two interesting features -- first, it correctly identifies the linear $m \, c^2$ offset from~\refeq{BFitSN}, and this provides an estimate of the error in the method -- we should have a value of $1.0$ in front of the $\bar m$ term in~\refeq{BfitSNC}, but instead we get $1.00123$, an error of $\approx .1\%$.  As for the coefficient of the $\bar m^5$ term, it agrees well with the accepted value of $-.163$~\cite{ Harrison,  Bernstein}.

\subsection{Self-consistent, Self-Coupled}

\begin{figure}[htbp] 
   \centering
   \includegraphics[width=4in]{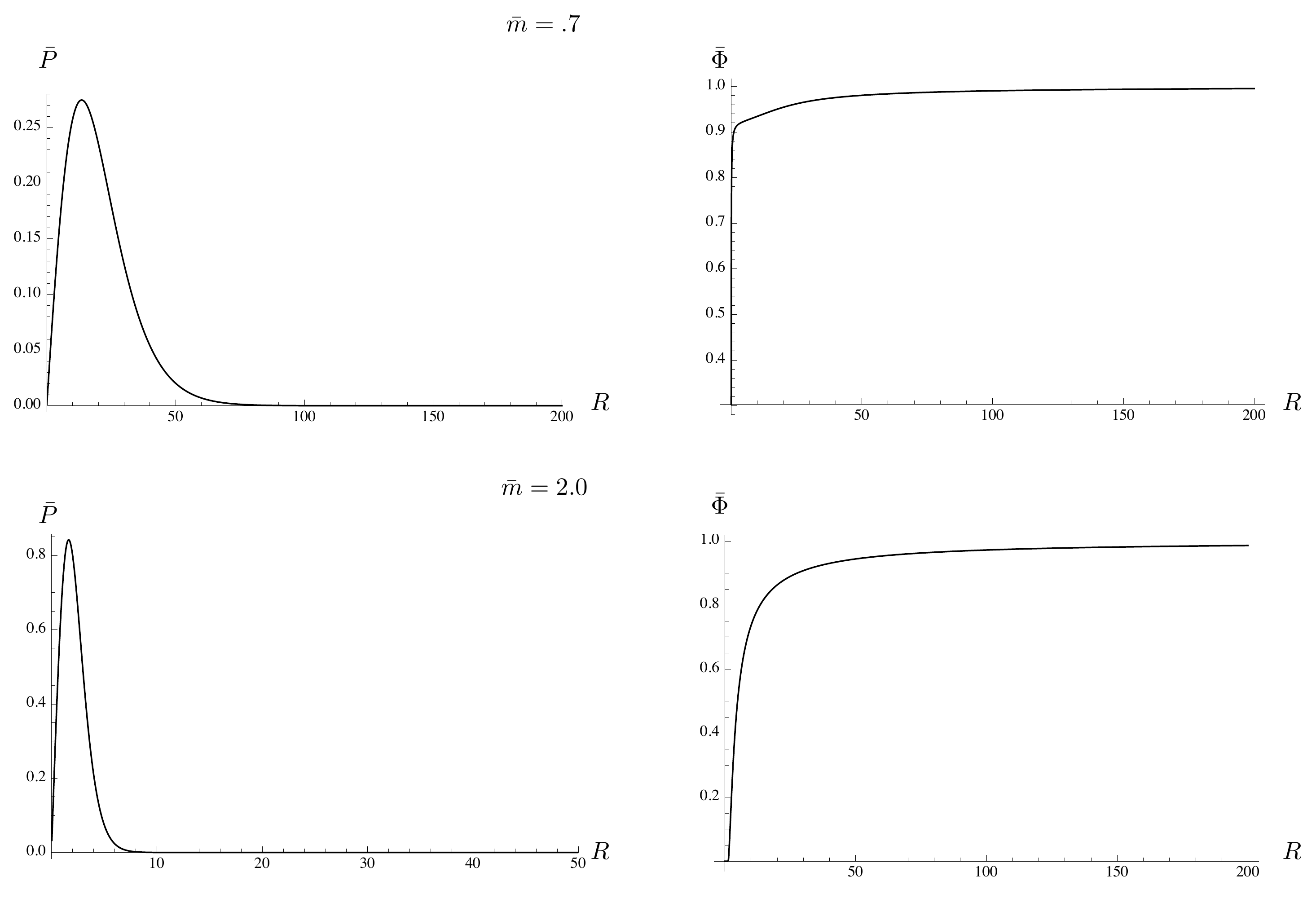} 
  \caption{The numerical solutions for $\bar P$ and $\bar\Phi$ for $\bar m = .7$ (top) and $\bar m = 2$ (bottom) for the self-coupled system.  Here, $R_\infty = 200$ with $\Delta R =200/5001$, and we use $\epsilon = 10^{-8}$ as the tolerance for the iteration (see~\refeq{stopper}).}
   \label{fig:SCSCPPhi}
\end{figure}

We use the same setup and parameters to find the ground state energies for the modified, self-sourcing gravity system.  In~\reffig{fig:SCSCPPhi}, we have the plots corresponding to~\reffig{fig:SNPPhi}, showing the functions $\bar P$ and $\bar\Phi$ for the smallest and largest mass values.

\begin{figure}[htbp] 
   \centering
   \includegraphics[width=4in]{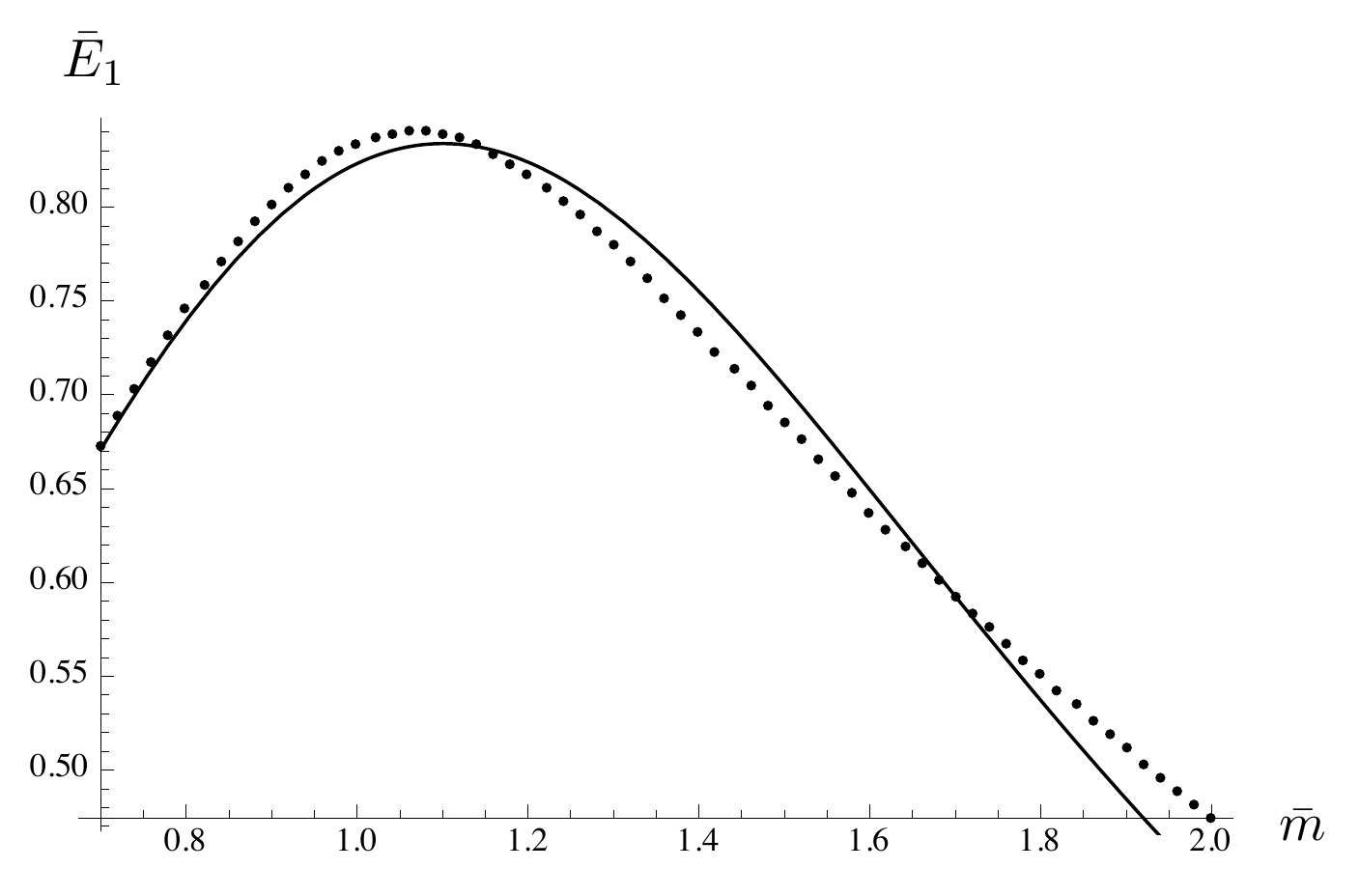} 
   \caption{Ground state energy as a function of mass for self-coupled gravity -- the points are numerically determined using the numerical method described in the previous section, and the curve is the best fit curve, $C(\bar m$), from~\refeq{BfitSCSCC}.}
   \label{fig:E1SCSC}
\end{figure}

This time, we fit to the function: $A \, \frac{2 \, \bar m}{2 + B \, \bar m^4}$, guided by the form of~\refeq{BFitSCSC}, and obtain
\begin{equation}\label{BfitSCSCC}
C(\bar m) = \frac{2.0205\, \bar m}{2 + 0.45448\, \bar m^4}.
\end{equation}
The numerically-determined ground state values and best fit curve are shown in~\reffig{fig:E1SCSC}.
While the fit curve captures the basic behavior of the ground state energies in this case, it does not fit as well as in the SN case.  The freedom in generating energies using $M_p$ in addition to $m$ allows for a more complicated spectrum (for SN, $m^5$ is the only possibility), and we do not expect the simple estimate from the Bohr method to work as well.  Still, the structure of the spectrum is described by that estimate.

\subsection{Comparison}

\begin{figure}[htbp] 
   \centering
   \includegraphics[width=4in]{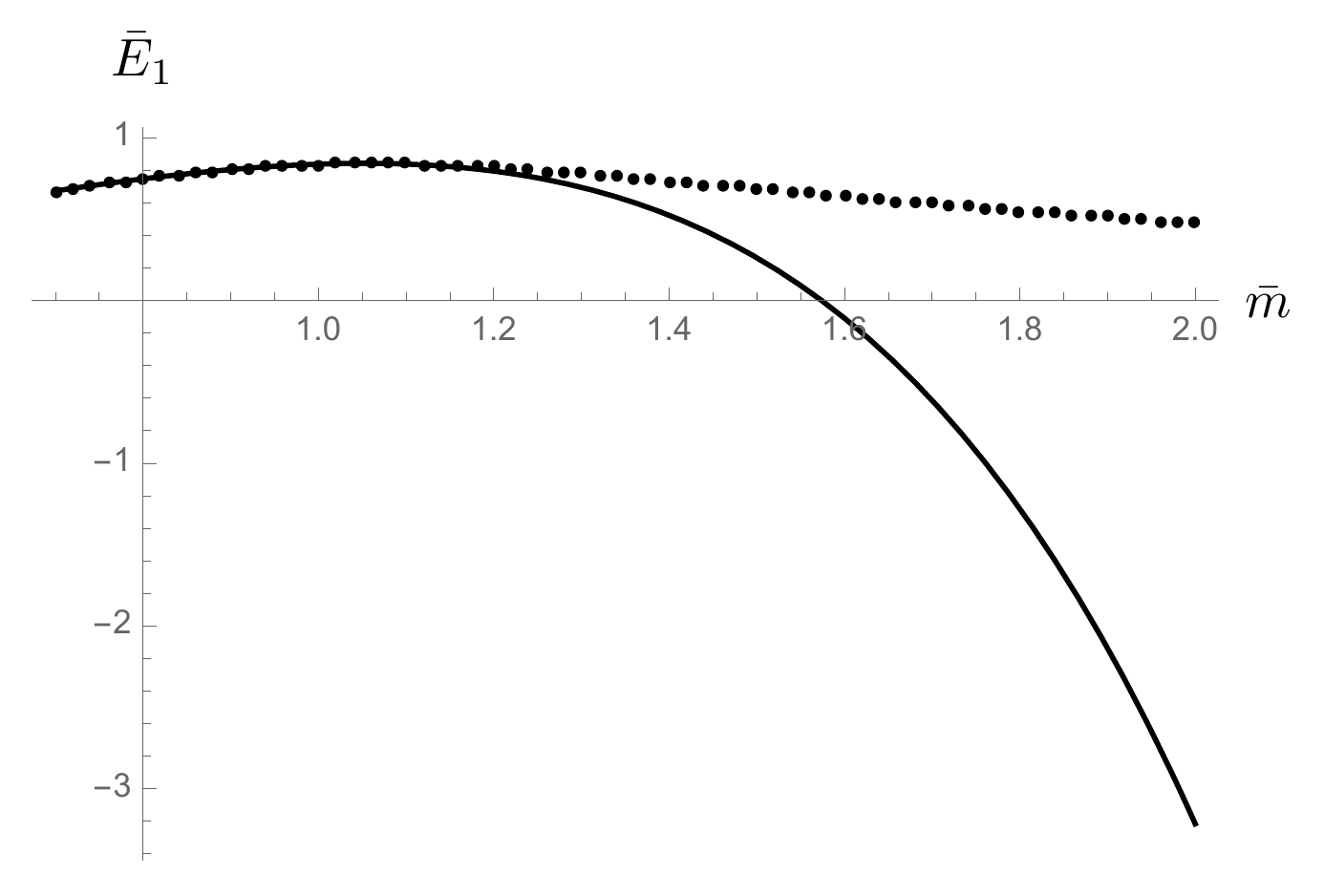} 
   \caption{Points show the ground state energy for the self-coupled system, while the solid line is the SN curve for the ground state energy.}
   \label{fig:SNSCE}
\end{figure}

We have the numerical spectrum from the self-coupled case, and can compare that with the $\bar m -.163 \, \bar m^5$ scaling from the SN system -- that is shown in~\reffig{fig:SNSCE} -- for small masses, the two ground state energies agree well, but they begin to diverge near $\bar m = 1.2$.  Note that both energies are offset by the same amount (these are bound state energies, so we expect them to be negative, they have been shifted upwards by the constant factor $m \, c^2$, the value of the potential energy at infinity).  The energies are off by around $8 \%$ by $\bar m = 2$.
While the SN energies become arbitrarily negative as $\bar m$ increases, the self-coupled spectrum looks like it asymptotically approaches zero (or $-m \, c^2$, relative to the value of the potential at spatial infinity) -- indeed, this is predicted by the Bohr estimate~\refeq{BohrSC}, where the energy scales like $1/m^3$ for $m$ large.

Another way to compare the spectrum of the self-coupled case with the SN energies is to look at the energy emitted during a transition from the first excited state to the ground state.  In the SN case, the first excited state has energy~\cite{Bernstein,Harrison}: $E_2 = m \, c^2 - .0308 \, G^2 \, m^5/\hbar^2$, and so the difference between the first excited state and the ground state is:
\begin{equation}
\Delta_{\hbox{\tiny{SN}}} = .1322 \, \frac{G^2 \, m^5}{\hbar^2},
\end{equation}
or $\Delta_{\hbox{\tiny{SN}}} = .1322 \, \bar m^5$ in our dimensionless variables.

\begin{figure}[htbp] 
   \centering
   \includegraphics[width=4in]{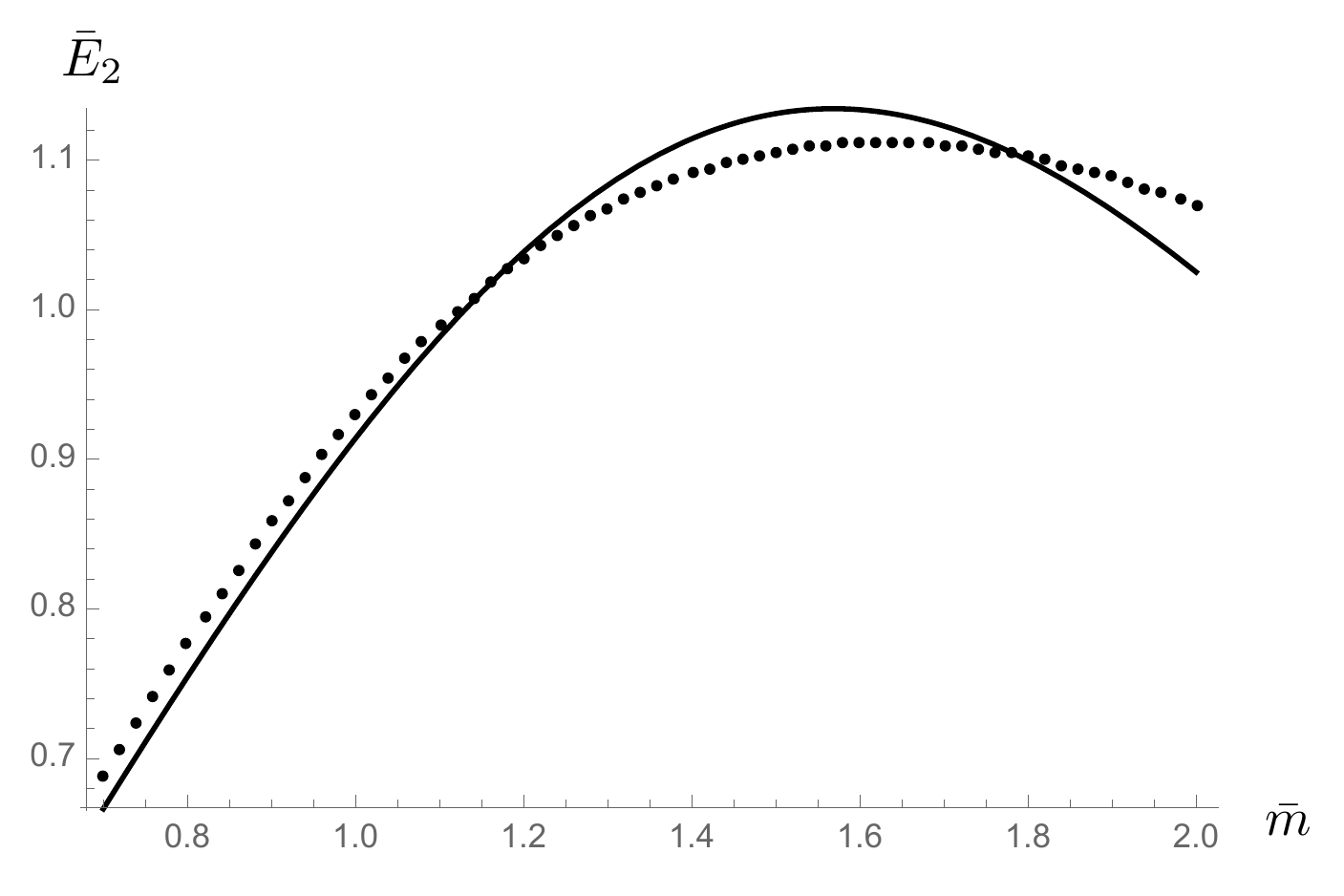} 
   \caption{The energy of the first excited state for the self-coupled case, together with its best-fit curve from the Bohr estimate.}
   \label{fig:SCSCFES}
\end{figure}

To compute the first excited state using our approach, we simply perform our iteration using the eigenvector associated with the second smallest eigenvalue (a simple modification of~\refeq{smalleval})-- the energy, as a function of $\bar m$ together with the best fit from~\refeq{BFitSCSC} with $n = 2$ (so that the fit function is $A \, \bar m/(8 + B \, \bar m^4)$) is shown in~\reffig{fig:SCSCFES}.

\begin{figure}[htbp] 
   \centering
   \includegraphics[width=4in]{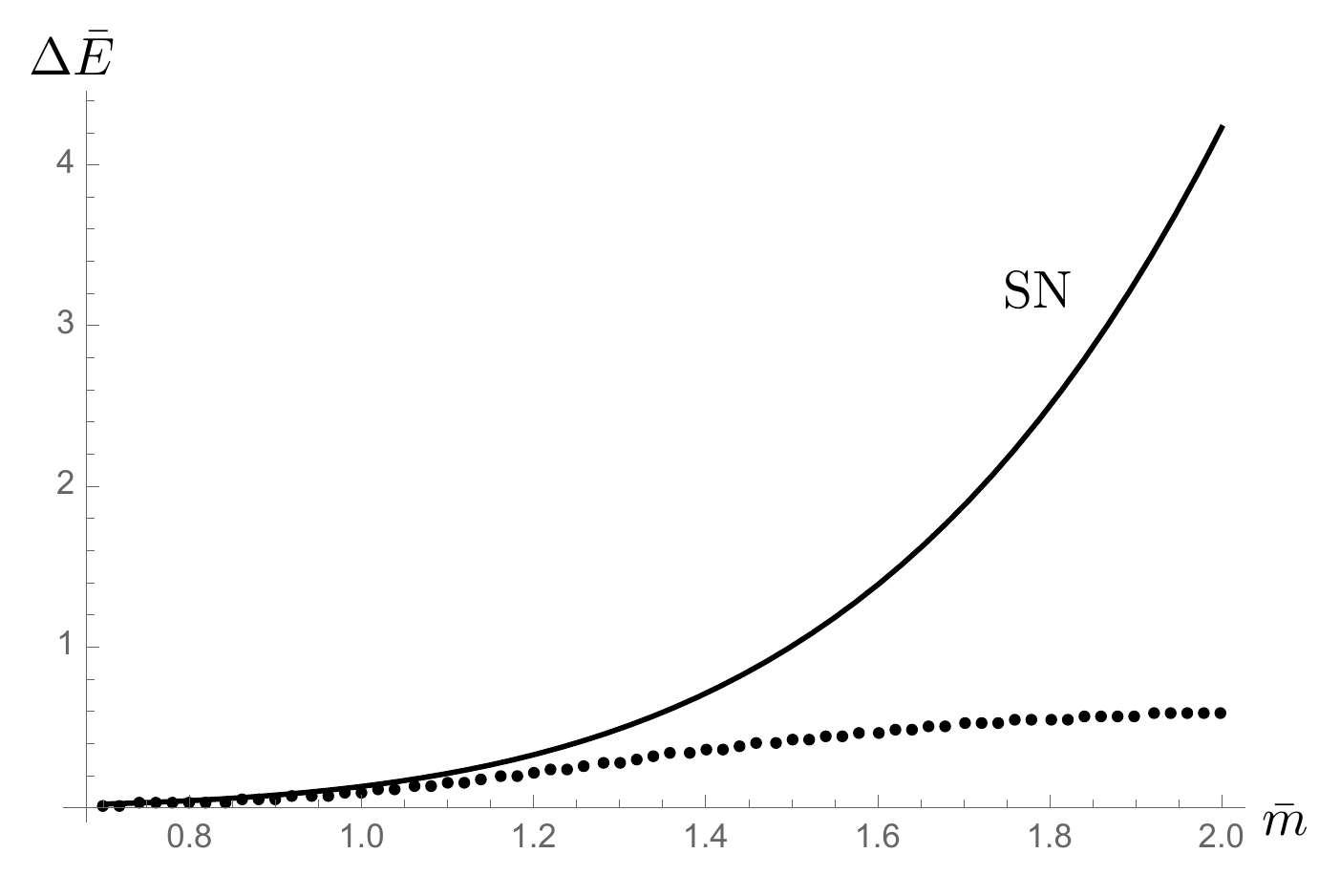} 
   \caption{The energy difference between the first excited state and ground state for SN (solid curve) and the self-coupled case (points).}
   \label{fig:SCSCDELTA}
\end{figure}

Subtracting the ground state for each mass, we get the $\Delta_{\hbox{\tiny SC}}$ to compare with the SN case -- that difference is shown in~\reffig{fig:SCSCDELTA}, together with the $\Delta_{\hbox{\tiny{SN}}}$ from above.
Notice there that by $\bar m = 2$, the energy difference for SN is roughly ten times that of the self-coupled case -- the frequency of emission (if radiation were emitted) for SN at this mass value would be ten times the frequency in the self-coupled case, providing a clear signature for one over the other.  If~\refeq{BohrSC} continues to hold qualitatively for larger $n$, we see that the self-coupled case has a maximum energy of $m\, c^2$ (the limit as $n \rightarrow \infty$), so any transition energy is bounded in this case, while the transition energies of SN will always be of the form: $\alpha \, \bar m^5$ for some constant $\alpha$.

\subsection{Total Energy in a Boson Collapse Model}

In the original application of the SN equations, appearing in~\cite{Bon}, the total energy of $N$ bosons in their ground state was calculated -- the Schr\"odinger-Newton pair appears as
\begin{equation}
\begin{aligned}
 i \, \hbar \, \frac{\dd \Psi(\vecbf r,t)}{\dd t} &= -\frac{\hbar^2}{2 \, m} \nabla^2 \Psi(\vecbf r,t) + m\, \Phi(\vecbf r, t) \, \Psi(\vecbf r,t) \\
\nabla^2 \Phi(\vecbf r, t) &= 4 \, \pi \, G \, N\, m \, \Psi^*(\vecbf r, t) \, \Psi(\vecbf r, t)
\end{aligned}
\end{equation}
where $N$ particles of mass $m$ are interacting gravitationally in the same state.  This is a very different application of the SN system (as compared with single-particle collapse), and yet we can use our self-consistent scalar gravity in place of the gravitational field equation here, just as for the single particle case.  We'll replace the Poisson equation with
\begin{equation}\label{Ninplace}
\nabla^2 \Phi(\vecbf r) = \frac{4 \, \pi \, G }{c^2} \,  N \, m\, \Psi^*(\vecbf r, t) \, \Psi(\vecbf r, t) \, \Phi(\vecbf r) + \frac{1}{2 \, \Phi(\vecbf r)} \, \nabla \Phi(\vecbf r) \cdot \nabla \Phi(\vecbf r).
\end{equation}
as usual.

\begin{figure}[htbp] 
   \centering
   \includegraphics[width=4in]{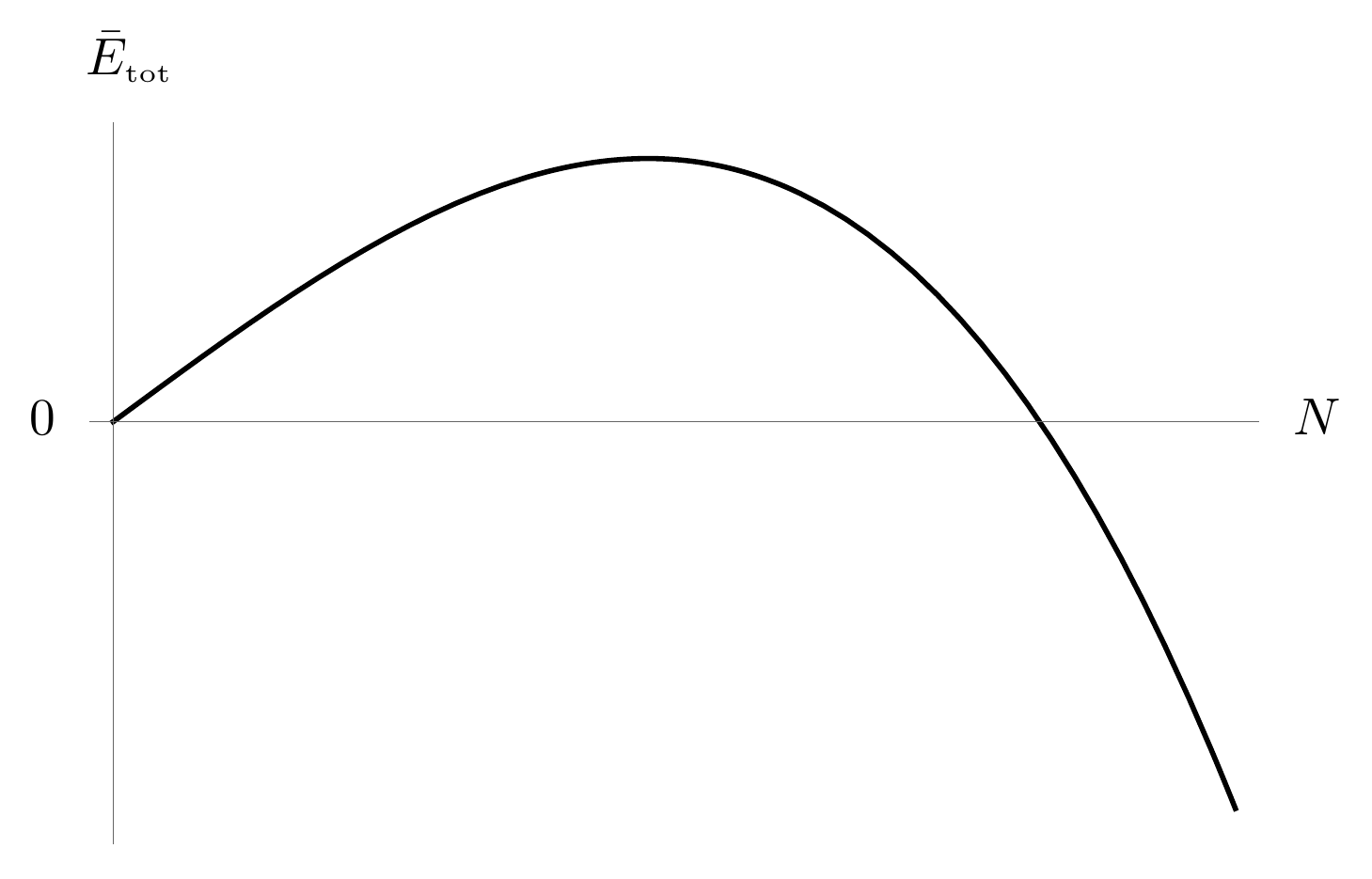} 
   \caption{The total energy, from~\cite{Bon}, for SN applied to a system of $N$ self-gravitating bosons in the ground state.}
   \label{fig:bonEtot}
\end{figure}

The total energy, from the SN approach, is $E_{\hbox{\tiny{tot}}} = N \, m \, c^2 - .163 \, N^3 \, G^2 \, m^5/\hbar^2$, a dimensionless energy of $\bar E_{\hbox{\tiny{tot}}} = \bar m\, N - .163 \, \bar m^5 \, N^3$ -- that is plotted in~\reffig{fig:bonEtot} (the choice of $m$ just changes the scale in $N$, so we have left an unscaled axis there).  There are three distinct regions in the total energy curve -- in the first, where the derivative of the total energy is positive, adding particles adds energy.  In the second, where the derivative is negative, but $\bar E_{\hbox{\tiny{tot}}}$ is still positive, adding particles decreases the total energy.  Finally, the total energy becomes negative when using the SN equations.

\begin{figure}[htbp] 
   \centering
   \includegraphics[width=4in]{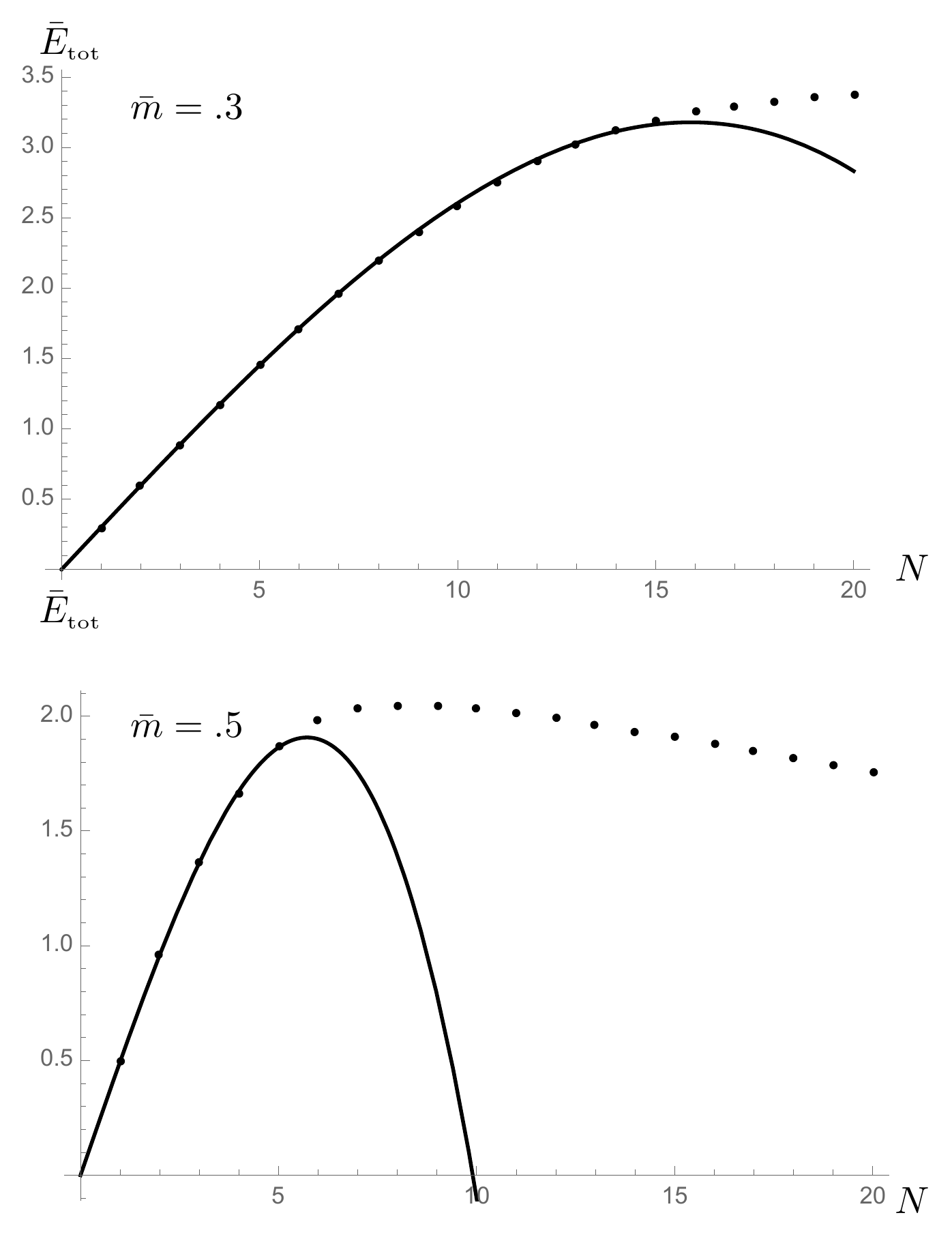} 
   \caption{Total energy for $N$ bosons, each with mass $\bar m = .3$ (top plot) or $\bar m = .5$ (bottom plot).  The points represent the total energy as calculated using self-consistent scalar gravity, while the solid curve is the SN prediction from~\cite{Bon}.}
   \label{fig:BOSTOTS}
\end{figure}

We can calculate the corresponding total energy  using the self-consistent scalar gravity in the form~\refeq{Ninplace} -- taking $\bar m = .3$ and $\bar m = .5$, we find the total energy and plot them with the SN result at those masses in~\reffig{fig:BOSTOTS}.  
The total energies are different between the two cases -- the peak that separates the region in which the total energy grows with additional particles from the region in which the total energy decreases with additional particles has shifted (upward in $N$ for the self-coupled case), and the self-coupled energy does not become negative.  There is additional physics which must be introduced (and is considered in detail in~\cite{Bon}) as the regimes change, and our goal here is only to once again draw a distinction (in principle detectable) between using the Poisson form of Newtonian gravity and using the nonlinear, self-coupled scalar gravity.

\section{Conclusion}

We have introduced an iterative numerical method to find the ground state energies for coupled quantum mechanical/gravitational problems.  We tested the method with the familiar Schr\"odinger-Newton pair of equations, and found that the method worked well, agreeing with known results for that system.  In addition, we used the Bohr method to predict the form of the ground-state energy for SN as a function of mass.  Then, we used the same approach for the self-consistent, self-sourced scalar gravity from~\cite{GiuliniSC}.  The spectrum has a very different mass dependence there, and this is to be expected given the additional (and necessary, for a theory of gravity) nonlinearity appearing in the gravitational field equation.  Once again, the Bohr method provided a qualitatively relevant estimate for the energy dependence on mass for both the ground state and first excited state.  The difference in energies provides a distinct signature for the self-coupled gravity -- the transition energy from the first excited state to the ground state is much less in the self-coupled case than in SN.

It would be interesting to explore the dynamics of the new, self-coupled case, and compare with the dynamics of SN, which are known.  Aside from specific computational targets, we use special relativity to motivate the self-sourcing in the gravitational field equation, but we use the non-relativistic Schr\"odinger equation to capture the quantum mechanical behavior -- we could probe the high energy solutions by using the Klein-Gordon or Dirac equations.  In these cases, it would be interesting to see how the self-coupled gravitational field arises in the context of~\cite{Giulini2} in which the authors show that Schr\"odinger-Newton is a natural limit of fields coupled to gravity, and the Poisson equation for the gravitational field appears as the first term in their expansion.  

\subsection*{Acknowledgements}
The authors thank David Griffiths for useful commentary and feedback.

\end{document}